# Gharibi_FaceCard for Contacts and Easy Personal - Information Exchange

**Wajeb Gharibi**[1] and **Gharib Gharibi**[2]

*Abstract – In this paper, we discuss a new contact way for exchanging personal information using mobile phones. The idea of this invention depends on allocating a special code (GC code) for each personal mobile, then creating files of information (GFC). When you request someone's GC code, the other party's phone will send you GFC; all specified personal data of that party, managed and saved previously. We think that this approach will facilitate the process of communication and exchanging the specified personal data easily, especially when acquaintance. Mobile number, phone and e-mail, for example, will be sent in a few seconds and using a simple code that does not exceed 6 characters long to transfer a huge amount of personal data through mobile devices rather than using traditional business /visiting cards.*

## I. INTRODUCTION

Although there are a lot of developments in the modern communication devices and techniques used universally by most of people, we have noticed that there are significant weaknesses and deficiencies in the method used for the exchange of the personal data.

The problem lies in that when you meet a new person, you often ask about his phone number, e-mail, his personal correspondence Skype, Facebook , or any other modern available communication tubes. Most of the used traditional methods are to write down the mobile number, email, or any other data, then saving it on the mobile phone. Or using business/visiting cards. Imagine that you will keep dozens of cards everywhere you go! We note that these cards are still in use despite of all developments of modern technologies and means of global communications! We can say that such methods are very poor and insufficient in their purpose for several reasons, including the following ones:

   a) Dictating of personal data using oral method is a traditional way. It needs a lot of time and efforts, but unfortunately, this method is still widely used despite of the existence of modern technologies.
   b) Carrying Business cards is annoying! One has to carry at least 10 cards for each attended incident.
   c) The information contained in these cards is limited and does not exceed the person's name, phone number and the organizations that the person works for.
   d) To exchange any personal information on these cards, one has to print a large number of business cards again.
   e) The design and printing of these cards require time, effort, regardless of cost.
   e) These cards can be quickly folded, damaged by water or humidity, regardless of losing them.

We have faced all of these problems and many others when we exchanged business cards with others. We had to find a new advanced technological way to contact them after the loss or misplace these cards or forgetting the name of the person who we are looking for. Sometimes we forget where we met that person which his mobile number is saved in our phones! So, we developed a new idea using the modern technology (mobile phones) owned by each person to facilitate the exchange of personal information and adding new contacting method to mobile phones.

Before explaining the suggested technique, it would be better to mention that there are many communication methods handling the same problem, but we have not found anyone yet uses any of those methods [1-4].

The rest of our paper is organized as follows. Section 2 describes the proposed technique. Section 3 defines the GC code. Conclusion is given in Section 4.

## II. GENERAL DESCRIPTION OF OUR TECHNIQUE

The main objective of our suggested technique is to exchange quickly and easily the personal data such as mobile numbers, emails, facebook, twitter, and skype accounts, etc. Rather than follow the inadequate traditional method. We propose a new communication technique to exchange huge amount of personal data using a simple-short code in a very short time and satisfied manner.

It is just a software application for mobile phones that allows users to enter and save all of their personal data in multiple files according to their needs. When somebody asks you about your phone number or email address or any other information, then you just need to tell him your mobile GC code, which is set and defined before. As soon as he enters your GC code in his mobile, he will receive all information you have saved, managed and intended to send him earlier. The proposed exchanging method could be done via Bluetooth to send data for short distances or via communication satellite service when long distances.

## III. GC CODE

GC is a code that assigned to each mobile phone by the user himself and authenticated by the user phone number. This code will make the exchanging process of personal information easily and fast without the need of sending these information by the owner himself, since all saved personal data will be sent automatically to those who requested GC code. So, the owner can send his personal information to hundreds of people in one room simultaneously! GC code consists of 2 up to 6 characters long and can contain alphabets, digits or any other symbols which are available in smart mobile phones.

[1] *Department of Computer Engineering & Networks, Jazan University, Jazan, Saudi Arabia.*

[2] *School of Computing and Engineering, Univ. of Missouri-Kansas City, USA.*
*E-mail:* [1]*Gharibi@jazanu.edu.sa and* [2] *ggk89@mail.umkc.edu*

Every person, by the help of our suggested technique, can have on his mobile phone, a great database of information of those people he has met in his life.

## IV. CONCLUSION

In this paper, we explained a new technique for exchanging personal data easily. This can be very helpful for Businessmen, Researchers, Students, Lecturers, and Professionals to send and receive large amounts of information about their operations using a simple code. Our application could deal with multimedia files as a future work.

## APPENDIX

Here, we give a brief explanation of our proposed technique depicted by figures.

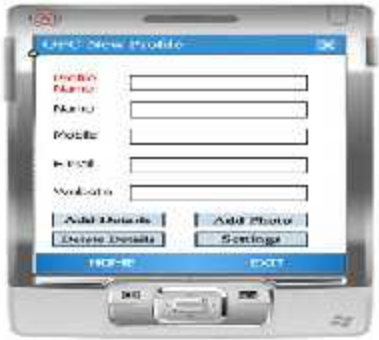

**Figure 1: Graphical user interface.**

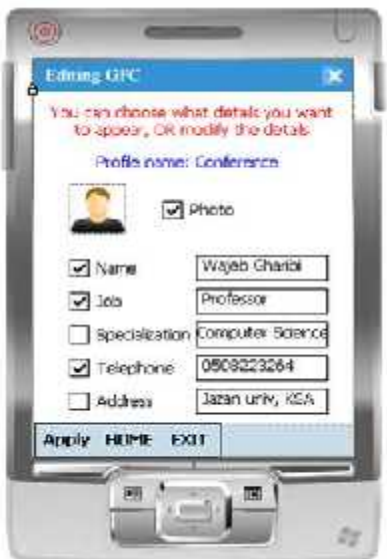

**Figure 2: Modified software interface**

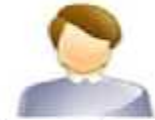

Hello. May I have you phone number, please

**Figure 3: Asking for exchanging personal data**

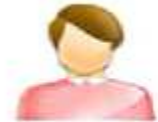

Of course. Here is my GC: Wa10

**Figure 4: Giving GC code.**

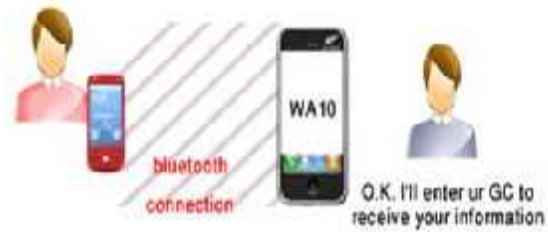

O.K. I'll enter ur GC to receive your information

**Figure 5: Sending and receiving information.**

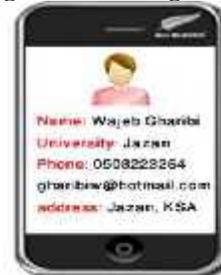
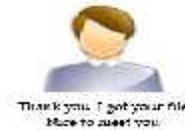

Thank you. I got your file Nice to meet you.

**Figure 6: Having received the needed information**

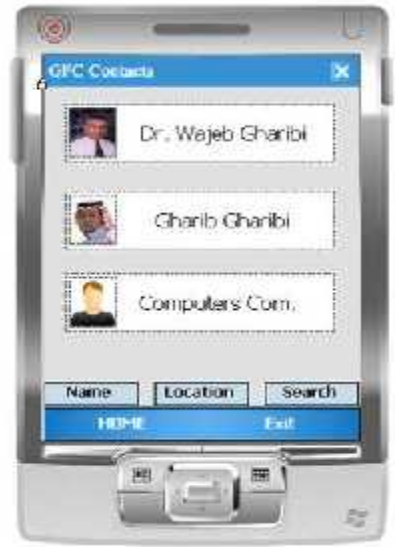

**Figure 7: Saving classified information.**